\documentclass[prl,twocolumn]{revtex4-1}
\usepackage{graphicx}
\usepackage{subfigure}
\usepackage{color}
\usepackage{amsmath}
\usepackage{amssymb}
\usepackage{wasysym}
\usepackage{bm}
\usepackage{placeins}
\usepackage[bookmarks=false]{hyperref}
\usepackage[usenames,dvipsnames]{xcolor}

\newcommand{\be}{\begin{equation}}
\newcommand{\ee}{\end{equation}}

\newcommand{\wo}{\omega_0}
\newcommand{\Ef}{E_F}
\newcommand{\epsk}[1]{\epsilon_{#1}}

\newcommand{\Xsc}{\chi_{sc}}
\newcommand{\Xcdw}{\chi_{cdw}}

\begin{document}

\title{Breakdown of Migdal-Eliashberg theory; a determinant quantum Monte Carlo study}
\author{I.~Esterlis$^1$, B.~Nosarzewski$^{1,2}$, E.~W.~Huang$^{1,2}$, B.~Moritz$^2$, T.~P.~Devereaux$^{2,3}$, D.~J.~Scalapino$^4$, S.~A.~Kivelson$^{1,3}$}
\affiliation{$^1$Department of Physics, Stanford University, Stanford, California 94305, USA \\
$^2$Stanford Institute for Materials and Energy Sciences, SLAC National Accelerator Laboratory and Stanford University, Menlo Park, CA 94025, USA \\
$^3$Geballe Laboratory for Advanced Materials, Stanford University, Stanford, CA 94305, USA \\
$^4$Department of Physics, University of California, Santa Barbara, CA 93106-9530, USA
}

%\date{\today}

\begin{abstract}
The superconducting (SC) and charge-density-wave (CDW) susceptibilities of the two dimensional Holstein model are computed using determinant quantum Monte Carlo (DQMC), and compared with results computed using the Migdal-Eliashberg (ME) approach. We access temperatures as low as 25 times less than the Fermi energy, $\Ef$, which are still above the SC transition. We find that the SC susceptibility at low $T$ agrees quantitatively with the ME theory up to a dimensionless electron-phonon coupling $\lambda_0 \approx 0.4$ but deviates dramatically for larger $\lambda_0$. We find that for large $\lambda_0$ and small phonon frequency $\wo \ll \Ef$ CDW ordering is favored and the preferred CDW ordering vector is uncorrelated with any obvious feature of the Fermi surface.
\end{abstract}

\maketitle

\emph{Introduction:} 
The electron-phonon (e-p) problem is of broad importance in solid-state physics, and especially so in the theory of superconductivity (SC). 
In this context, a key question is what are the conditions that lead to the highest possible SC transition temperature, $T_c$.
Given that this occurs when the dimensionless e-p coupling $\lambda_0$ is not small, 
one would a priori expect  this question might be  analytically unanswerable.  However, Migdal-Eliashberg (ME) theory purports to be valid  even if $\lambda_0$ is not small, provided $\lambda_0 \wo/\Ef$ is small, where $\wo$ is an average phonon frequency and $\Ef$ is the Fermi energy \cite{migdal1958interaction,eliashberg1960interaction}. On the other hand, from a strong-coupling (large $\lambda_0$) perspective,  it is clear ME theory breaks down for large $\lambda_0$ no matter how small $\wo/\Ef$, due to the formation of bipolarons \cite{0295-5075-56-1-092,PhysRevB.84.184531,Carlson2008}. Thus, one faces the practical question: at what value of the e-p coupling does ME theory break down, and how? 

\emph{Model:} 
To be  explicit, we  consider the two-dimensional Holstein Hamiltonian \cite{HOLSTEIN1959325}
	\be
	H = H_e + H_p + H_{ep},
	\label{eq:Hamiltonian}
	\ee
where
	\be
	\begin{aligned}
	H_e &= -\sum_{ij,\sigma}t_{ij}(c^\dag_{i\sigma}c_{j\sigma} + \mathrm{h.c.}) - \mu\sum_{i,\sigma}n_{i\sigma}, \\
	H_p &= \sum_i \left(\frac{P_i^2}{2M} + \frac 12 K X_i^2\right), \\
	H_{ep} &= \alpha\sum_{i,\sigma} n_{i,\sigma} X_i,\\
	\end{aligned}
	\ee 
$c^\dag_{i\sigma}$ creates an electron on site $i$ with spin polarization $\sigma = \uparrow,\downarrow$, $n_{i\sigma}=c^\dag_{i\sigma}c_{i\sigma}$ 
is the local electronic density,  $P_i$ and $X_i$ are the position and momentum operators of Einstein phonons with mass $M$, and $\mu$ is the chemical potential.   The bare phonon frequency is thus $\wo = \sqrt{K/M}$, and $\alpha$ is the e-p coupling constant. (We take units in which $M=\hbar = k_B = 1$.)
 
There are two important dimensionless parameters in the model: 
the adiabatic parameter $\wo/\Ef$ and the dimensionless e-p coupling
	\be
	\lambda_0 = \alpha^2\rho(\Ef)/K,
	\label{eq:lam_0}
	\ee
where $\rho(\Ef)$ is the density of states at the Fermi energy. To make contact with other approaches we also present data as a function of a ``renormalized" coupling, denoted by $\lambda$, which we define in analogy with the phenomenological coupling extracted from tunneling spectra and often used in studies of ``strongly-coupled" superconductors \cite{ALLEN19831}. In the limit of weak coupling,  $\lambda=\lambda_0 + {\cal O}(\lambda_0^2)$, but for $\lambda_0 \sim 1$,  we will see that phonon softening leads to $\lambda > \lambda_0$. The prescription for computing  $\lambda$ will be explained below in Eq.\eqref{eq:lam_phen}.

We investigate this model numerically via determinant quantum Monte Carlo (DQMC) simulations and analytically via ME theory. Details of the DQMC algorithm, including explanation of both the local and global phonon field updates used, can be found in \cite{johnston2013determinant}. Unless stated otherwise,  we work with a square lattice with both nearest-neighbor and next-nearest-neighbor hopping $t'/t=-0.3$ and a fixed density $n=0.8$. We keep a nonzero $t'$ to avoid nesting near half-filling and also because previous studies have found that nonzero $t'$ leads to an enhanced pairing response \cite{PhysRevB.46.271} .  We have studied systems of linear size $L=8-12$ with periodic boundary conditions and temperatures $T = \beta^{-1} = t/4$ to $t/16$. All data in the main text is shown for our largest system size $L=12$, which is large enough that most observables are essentially $L$ independent, i.e. are characteristic of the thermodynamic limit. The DQMC results are shown as  solid symbols  in the various figures and where error bars are not visible, the statistical error is less than the symbol size. The figures also show comparisons of the DQMC results with ME theory, which is shown as either continuous curves or open symbols. ME calculations have also been carried out on system size $L=12$. All data in the main text has the adiabatic ratio $\wo/\Ef = 0.1$, which puts us comfortably within the putative regime of validity of ME theory. In the Supplemental Material we present data for other values of $\wo/\Ef$. We note that this model is free of the notorious minus-sign problem and hence we are able to access relatively large system sizes and low temperatures. However, for the parameters used here, we are still unable to access temperatures $T \lesssim t/16$ due to prohibitively long phonon autocorrelation times \cite{hohenadler2008autocorrelations}.

\emph{DQMC Results:}
While we typically cannot access sufficiently low temperatures to observe transitions to either a SC or a charge-density wave (CDW) phase, we do access low enough $T$ that a significant growth of the corresponding susceptibilities can be measured, showing the ordering tendencies of the system.  The $s$-wave pair susceptibility  is defined as
	\be
	\Xsc = \int_0^\beta d \tau ~ \langle \Delta(\tau)\Delta^\dag(0)\rangle,
	\label{eq:Xsc_def}
	\ee
where
	\be
	\Delta^\dag = \frac 1L \sum_i c^\dag_{i\uparrow}c^\dag_{i\downarrow}.
	\ee
and the CDW susceptibility is
	\be
	\Xcdw(\mathbf q) =  \int_0^\beta d\tau ~ \langle \rho_\mathbf q(\tau)\rho^\dag_\mathbf q(0)\rangle,
	\ee
where 
	\be
	\rho_\mathbf q^\dag = \frac 1L\sum_{i,\sigma}e^{i\mathbf q\cdot \mathbf R_i} c_{i\sigma}^\dag c_{i\sigma}.
	\label{eq:Xcdw_def}
	\ee
We will use the symbol $\Xcdw$ to represent the value of $\Xcdw(\mathbf q)$ evaluated at  $\mathbf q \equiv \mathbf Q_{max}$ at which it is maximal.

%%%%%%%%%%%%%%%%%%%%%%%%%%%%%%%%%%%%%%%%%%%
\begin{figure}[t]
  \centering
 \includegraphics[]{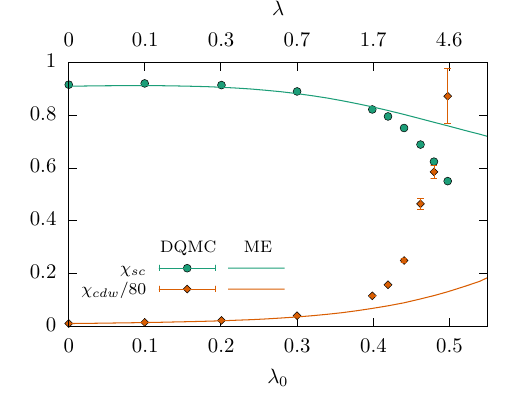}
 \caption{$\chi_{sc}$ and $\chi_{cdw}$ as a function of $\lambda_0$ (lower scale) and $\lambda$ (upper scale) for $\wo/\Ef=0.1$ at base temperature,  $\beta t= 16$,  density $n=0.8$, and $L=12$. Data points are DQMC values, solid lines are computed in ME approximation. We see a breakdown in the ME theory for $\lambda_0 \gtrsim 0.4$ ($\lambda \gtrsim 1.7$). }
 \label{fig:Xsc_vs_lambda}
\end{figure}
%%%%%%%%%%%%%%%%%%%%%%%%%%%%%%%%%%%%%%%%%%%

\begin{figure}[t]
  \centering
 \includegraphics[]{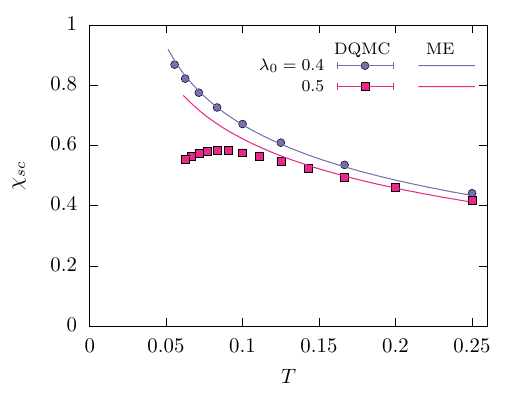}
 \caption{$\Xsc$ as a function of $T$ for fixed values of $\lambda_0 = 0.4$ and $0.5$ ($\lambda \approx 1.7$ and $4.6$), $\wo/\Ef=0.1$, $n=0.8$, and $L=12$. For $\lambda_0 = 0.4$, the ME theory accurately captures the behavior of $\Xsc$ over the entire temperature range while for $\lambda_0 = 0.5$ the theory is qualitatively incorrect.}
 \label{fig:Xsc_vs_T}
\end{figure}

\begin{figure}[t]
\includegraphics[]{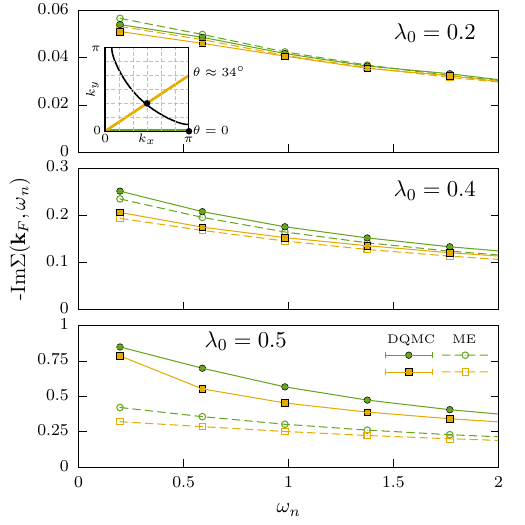}
\caption{Imaginary part of the electronic self energy, $\mathrm{Im}\Sigma(\mathbf k \approx \mathbf k_F,\omega_n)$, where $\omega_n = (2n+1)\pi T$ and $\mathbf k \approx \mathbf k_F$ is a momentum near the Fermi surface, evaluated at $\theta = 0$ and $\theta \approx  34^{\circ}$. The inset shows the $\mathbf k$-space mesh for an $L=12$ grid and the two points at which $\mathrm{Im}\Sigma$ is evaluated. These points correspond to the points closest to the zone diagonal and zone boundary respectively of the Fermi surface. The ME theory captures both the Matsubara frequency and momentum dependence of $\mathrm{Im}\Sigma$ for $\lambda_0 \leq 0.4$ but again shows a breakdown for $\lambda_0 = 0.5$. Other parameters are $\wo/\Ef=0.1$, $\beta t= 16$, $n=0.8$, and $L=12$.}
\label{fig:ImSig}
\end{figure}

In Figure \ref{fig:Xsc_vs_lambda} we plot $\Xsc$ and $\Xcdw$ as a function of $\lambda_0$ (lower axis) and $\lambda$ (upper axis) at the lowest studied temperature $\beta t = 16$. 
The results from the ME theory, discussed in more detail in the next section, agree quantitatively with the data  for $\lambda_0 \leq 0.4$ ($\lambda \lesssim 1.7$), after which the SC susceptibility takes a sharp downturn while the ME theory shows no similar features. We thus conclude that, for the parameters considered, ME theory breaks down dramatically for $\lambda_0 \gtrsim 0.4$.  This is consistent with dynamical mean-field theory (DMFT) results reported by Bauer et al. \cite{PhysRevB.84.184531}. Also evident from Fig.~\ref{fig:Xsc_vs_lambda} is that the downturn in $\Xsc$ is accompanied by a rapid rise of CDW correlations, which in turn leads to phonon softening and a corresponding increase in $\lambda/\lambda_0$. We find that $\Xcdw$ is peaked at wave-vector $\mathbf Q_{max} = (\pi,\pi)$ which, we emphasize, is a wave-vector not associated with any obvious features of the Fermi surface (see Inset of  Fig.~\ref{fig:ImSig} for the Fermi surface). The abrupt nature of the breakdown of ME theory can also be seen in Fig.~\ref{fig:Xsc_vs_T}, where we plot $\Xsc$ as a function of $T$ for $\lambda_0=0.4,0.5$ ($\lambda \approx 1.7,4.6$). For $\lambda_0 = 0.4$, ME theory shows good agreement with the data over the entire temperature range $\beta t = 4 - 16$. However, for $\lambda_0  = 0.5$, while the ME theory does predict a decrease in the pairing response relative to $\lambda_0 = 0.4$, it clearly misses even the qualitative behavior of $\Xsc$. 

%%%%%%%%%%%%%%%%%%%%%%%%%%%%%%%%%%%%%%%%%%%
%%%%%%%%%%%%%%%%%%%%%%%%%%%%%%%%%%%%%%%%%%%

We have also computed the electron and phonon imaginary time ordered Green's functions. In Figure \ref{fig:ImSig} we plot the imaginary part of the electronic self-energy $\mathrm{Im}\Sigma$ for $\lambda_0 = 0.2,0.4,$ and $0.5$ ($\lambda \approx 0.3, 1.7, 4.6$), as a function of Matsubara frequency $\omega_n = (2n+1)\pi T$ and for two momenta near the Fermi surface. For $\lambda_0 = 0.2$ the self-energy is nearly momentum independent and the Matsubara frequency dependence of $\mathrm{Im}\Sigma$ is captured accurately by ME theory. For $\lambda_0 = 0.4$ the self-energy develops weak momentum dependence and the dependence on both $\mathbf k$ and $\omega_n$ is again captured well by ME theory. For $\lambda_0 = 0.5$ the self-energy remains weakly momentum  dependent in both the ME and DQMC results but ME theory drastically underestimates the magnitude of the self-energy. %However, by $\lambda_0 = 0.5$, we see strong deviation from the ME theory, which remains weakly momentum dependent and predicts a larger self-energy for $\mathbf k$ at the zone boundary than along the zone diagonal. On the other hand, the DQMC results show strong momentum dependence and a much larger self-energy along the zone diagonal. 

\begin{figure}[t]
\includegraphics[]{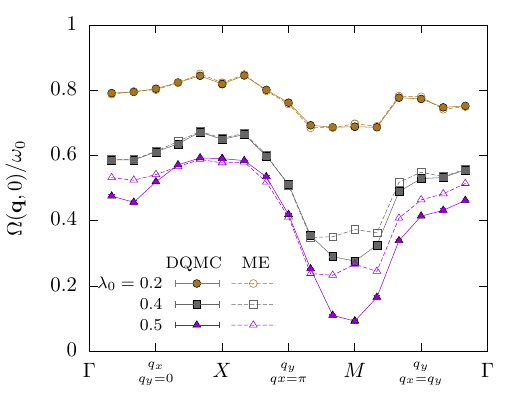}
\caption{Ratio of renormalized to bare phonon frequency. The renormalized phonon frequency is defined in Eq.\eqref{eq:D}. For $\lambda_0 \leq 0.4$ we see ME theory accurately predicts the momentum dependence of $\Omega(\mathbf q,0)$. However, for $\lambda_0 = 0.5$, ME theory dramatically underestimates the softeninng of the phonon propagator at $\mathbf q = (\pi,\pi)$. Other parameters are $\wo/\Ef=0.1$, $n=0.8$, $\beta t= 16$, and $L=12$.}
\label{fig:w0_renorm}
\end{figure}

In Figure \ref{fig:w0_renorm} we plot the renormalized phonon frequency $\Omega(\mathbf q, 0 ) = [\wo ^2 + \Pi(\mathbf q,0)]^{1/2}$, where $\Pi(\mathbf q,\nu_n)$ is defined implicitly in terms of the phonon Green's function according to
	\be
	D(\mathbf q,\nu_n) \equiv \frac{2\wo}{(i\nu_n)^2 - \wo^2 - \Pi(\mathbf q, \nu_n)}
	\label{eq:D}
	\ee
and $\nu_n = 2\pi n T$. For $\lambda_0 = 0.2$ and $0.4$ we see that the ME theory captures the renormalization of the phonon propagator with remarkable accuracy. (For $\lambda_0=0.4$, there is a noticeable error in the ME result in a narrow range of ${\mathbf q}$ around $(\pi,\pi)$;  this reflects an emerging problem in treating the CDW tendencies, as is discussed further in the Supplemental Material.) However, for $\lambda_0 = 0.5$, ME theory drastically underestimates (by a factor of $\sim 3$) the phonon softening at $\mathbf q = (\pi,\pi)$ and also gives a weaker softening near $\mathbf q = 0$.

%%%%%%%%%%%%%%%%%%%%%%%%%%%%%%%%%%%%%%%%%%%
%%%%%%%%%%%%%%%%%%%%%%%%%%%%%%%%%%%%%%%%%%%

%%%%%%%%%%%%%%%%%%%%%%%%%%%%%%%%%%%%%%%%%%%
%%%%%%%%%%%%%%%%%%%%%%%%%%%%%%%%%%%%%%%%%%%%

\begin{figure}[t]
  \centering
  \includegraphics[scale=0.5]{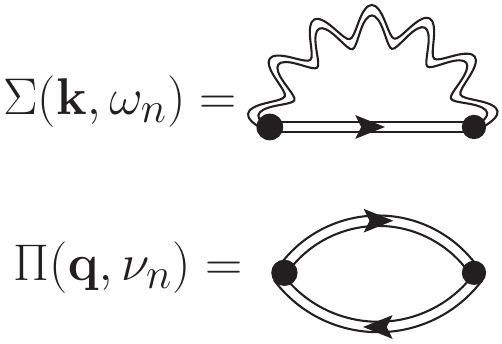}
\caption{Migdal equations for electronic and phonon self energies in the normal state. Double lines indicate fully renormalized Green's functions and the solid dot is the bare vertex $\alpha$.}
\label{fig:Migdal_Equations}
\end{figure}

\emph{Migdal-Eliashberg Theory:}
The ME theory for the normal state of an interacting e-p system can be summarized by the diagrams in Figure \ref{fig:Migdal_Equations}, which constitute a set of closed self-consistent equations for the electron and phonon self-energies. The approach is justified by the observation that the leading correction to the e-p vertex is proportional to $\wo/\Ef$, and hence can be ignored for $\wo/\Ef \ll 1$ \cite{migdal1958interaction,eliashberg1960interaction}. As pointed out in \cite{PhysRevB.42.2416}, it is important to include self-consistently the equation for the phonon self-energy (rather than just using the bare phonon propagator in the electron self-energy equation) to account for effects due to phonon softening near a CDW transition. Details of the numerical procedure used to solve these equations and how the self-energies are used to compute various observables can be found in, e.g., \cite{PhysRevB.42.2416,PhysRevB.40.197}.

An important quantity entering the ME theory is the coupling constant $\lambda$ \cite{ALLEN19831}, defined as 	
	\be
	\lambda = 2\int_0^\infty d\omega ~ \frac{\alpha^2F(\omega)}{\omega},
	\label{eq:lam_phen}
	\ee
where $\alpha^2 F(\omega) =  \rho(E_F) \alpha^2 \langle B(\mathbf k - \mathbf k',\omega) \rangle_{FS}$. Here $B(\mathbf q,\omega)$ is the phonon spectral function and the brackets denote the Fermi surface (FS) average%an appropriate Fermi surface average (see Supplemental Material). 
	\begin{align}
	&\langle B(\mathbf k - \mathbf k',\omega) \rangle_{FS}  = \frac{1}{\rho(E_F)^2} \\
	&\times\int \frac{d^2\mathbf k}{(2\pi)^2} \frac{d^2\mathbf k'}{(2\pi)^2} B(\mathbf k - \mathbf k',\omega) \delta(\epsk{\mathbf k} - \Ef)\delta(\epsk{\mathbf k'} - \Ef). \nonumber
	\end{align}
To extract this quantity from DQMC data we use the relationship between the imaginary time ordered phonon Green's function and the spectral function,
	\be
	D(\mathbf q,\nu_n) = \int_0^\infty d\omega ~ B(\mathbf q,\omega) \frac{2\omega}{(i\nu_n)^2 - \omega^2},
	\ee
from which it follows that
	\be
	\lambda = -\frac{\lambda_0 \omega_0}{2} \langle D(\mathbf k - \mathbf k',0) \rangle_{FS}.
	\ee
	
%%%%%%%%%%%%%%%%%%%%%%%%%%%%%%%%%%%%%%%%%%%%	
%%%%%%%%%%%%%%%%%%%%%%%%%%%%%%%%%%%%%%%%%%%%	
	
\emph{Conclusions:} 
Comparing the DQMC results on the Holstein model with ME theory, we find remarkably good quantitative agreement for e-p coupling less than a crossover value, $\lambda_0 \lesssim \lambda^\star =0.4$.  However, as $\lambda_0$ exceeds $\lambda^\star$, increasingly dramatic quantitative and qualitative differences develop. This is despite the fact that $\lambda(\bar\Omega/E_F)$ (the nominal control parameter for ME theory) is still small, and that the ME theory shows no sign that a crossover has occurred.  (Note that while $\lambda$ increases over $\lambda_0$, the average phonon frequency $\bar\Omega$ decreases so that at $\lambda_0 = 0.5$, $\lambda \bar \Omega/\Ef \approx 0.3$.) This crossover appears in some ways analogous to a first order transition -- it involves a change in the character of the low energy theory to one which will eventually (at larger $\lambda_0$) be governed by the strong-coupling physics of bipolarons, commensurate CDWs, and phase separation \cite{0295-5075-56-1-092,PhysRevB.84.184531,Carlson2008}.  

%In the context of the search for higher $T_c$ conventional superconductivity, we %find

Our results  are also interesting in the context of the  quest for higher $T_c$ %conventional 
 superconductors.  %On the basis of the following analysis, we conclude that the superconducting %tendencies seem to be
%T$_c$ is maximized for $\lambda_0 \sim \lambda^\star$ ($\lambda \sim 2$):  % on the fact that, f
For $\lambda_0 \lesssim \lambda^\star$, the measured $\chi_{sc}$ agrees well with ME theory in the accessible range of temperatures, and hence %we can
it is reasonable to use the ME expression as a way to extrapolate the DQMC results to lower $T$.  By this line of reasoning, we can use the value of $T_c$ computed within ME as a reliable estimate of the true $T_c$ for $\lambda_0$ in this range. %Since for  $\lambda_0 \lesssim \lambda^\star$ the
Since the ME $T_c$ is an increasing function of $\lambda_0$ %and hence 
 we conclude the same is true of the actual $T_c$, so long as $\lambda_0 \lesssim \lambda^\star$. On the other hand, for $\lambda_0 = 0.5$ (where ME theory no longer agrees with the DQMC results), $\chi_{sc}$ from DQMC is a decreasing function of decreasing temperature (see Fig.\ref{fig:Xsc_vs_T}), from which we conclude $T_c$ has been substantially suppressed and %may in fact be zero.
likely vanishes. This implies $T_c$ is optimized around $\lambda_0 \approx \lambda^\star$, %and
where from ME theory we %predict a 
estimate that the maximal $T_c$ is $T_c^{(max)} \approx 8 \times 10^{-2} \omega_0$. 
 
 The property of an optimal $T_c$ should be contrasted with conventional ME theory, which predicts a monotonically increasing $T_c$ as a function of $\lambda_0$ \cite{PhysRevB.12.905,KRESIN1987434}. Of course, it is likely that the precise value of $\lambda^\star$ is non-universal, and there may be ways to engineer the model to increase it further - say by suppressing the CDW and/or polaronic tendencies.  For instance, Pickett \cite{Pickett2006}  has discussed the possibility of using multiple quasi-2D Fermi surfaces to enhance $T_c$ and Werman and Berg \cite{PhysRevB.93.075109} have recently shown that in a particular large $N$ limit, in which the number of phonon modes is large compared to the number of fermionic modes, one can access the large $\lambda$ limit without polaronic effects. However,  as seen in $\mathrm H_3 \mathrm S$ \cite{PhysRevLett.21.1748,H3S_SC}, it is probably a more promising route to increase $T_c$  by increasing $\omega_0$ (the prefactor in $T_c$) keeping $\lambda_0\approx \lambda^\star$, rather than increasing $\lambda$ \cite{PhysRevB.93.224501,PhysRevLett.75.1158,PhysRevLett.90.167006}. (On the other hand, increasing $\wo$ makes the effects of bare Coulomb repulsion -- which are completely absent in our treatment -- more important.)

Finally, our results should be put in the  context of previous studies of the Holstein model. The competition between SC and CDW has been studied  via DQMC, albeit with a different focus and in a different parameter regime. (See \cite{PhysRevB.40.197, PhysRevLett.66.778, PhysRevB.42.2416, PhysRevB.46.271} and references therein.) The Holstein model has also been studied extensively via DMFT. (See \cite{PhysRevB.48.6302,PhysRevB.48.3881,PhysRevLett.89.196401,Hague2008,PhysRevB.84.184531,PhysRevB.58.14320,PhysRevLett.91.186405} for discussions of the crossover between weak and strong coupling as well as assessments of the validity of ME theory for the Holstein model.) The conclusions of these studies are broadly similar to those reached here.  However, because the DMFT is  done in infinite dimensions, we are unable to make quantitative comparisons with these studies. Using the dynamical cluster approximation (DCA), the inclusion of lowest-order vertex corrections has been studied \cite{0953-8984-15-17-309}. In that study it was found that inclusion of vertex corrections tends to return the system to the ME regime by averting phonon softening. As we have seen, however, the ME theory already underestimates the phonon softening, suggesting that the inclusion of vertex corrections will not save the theory. 

We would like to thank Akash Maharaj, Raghu Mahajan, and Abolhassan Vaezi for collaboration during early stages of this work. We acknowledge insightful discussions with Sri Raghu, Samuel Lederer, and Yoni Schattner.  IE, BN, EWH, BM, and TPD were supported by the U.S. Department of Energy, Office of Basic Energy Sciences, Division of Materials Sciences and Engineering, under Contract No. DE-AC02-76SF00515. SAK was supported by NSF DMR-1608055. DJS was supported by the Scientific Discovery through Advanced Computing (SciDAC) program funded by U.S. Department of Energy, Office of Science, Advanced Scientific Computing Research and Basic Energy Sciences, Division of Materials Sciences and Engineering. This research was supported in part by the National Science Foundation under Grant No. NSF PHY11-25915. Computational work was performed on the SIMES and Sherlock clusters at Stanford University.

\bibliographystyle{apsrev4-1} 
\bibliography{holstein+migdal}

\newpage

~

%%%%%%%%%% Merge with supplemental materials %%%%%%%%%%
%%%%%%%%%% Prefix a "S" to all equations, figures, tables and reset the counter %%%%%%%%%%
%\pagebreak
\newpage
\widetext
\begin{center}
\textbf{\large Supplementary Material: ``Breakdown of Migdal-Eliashberg theory; a Quantum Monte Carlo study''}
\end{center}
\setcounter{equation}{0}
\setcounter{figure}{0}
\setcounter{table}{0}
\setcounter{page}{1}

\renewcommand{\theequation}{S\arabic{equation}}
\renewcommand{\thefigure}{S\arabic{figure}}
\renewcommand{\bibnumfmt}[1]{[S#1]}
%\renewcommand{\citenumfont}[1]{S#1}
%%%%%%%%%% Prefix a "S" to all equations, figures, tables and reset the counter %%%%%%%%%%

%\title{Supplemental Material: Breakdown of Migdal-Eliashberg theory; a Quantum Monte Carlo study}
%\author{I. Esterlis, B. Nosarzewski, E. W. Huang, B. Moritz, T. P. Devereaux, D. J. Scalapino, S. A. Kivelson}
%\affiliation{}
%\date{\today}
%\maketitle

\section{CDW details}

Here we provide details on the charge-density-wave discussed in the main text. Parameters for all data discussed in this section are the same as those for data in the main text: nearest-neighbor and  next-nearest-neighbor hopping $t'/t = -0.3$, density $n = 0.8$, and adiabatic ratio $\omega_0 /\Ef= 0.1$. 

\begin{figure}[t]
  \centering
 \includegraphics[]{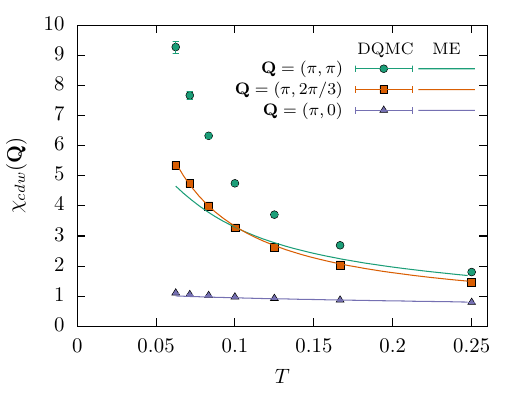}
 \caption{$\Xcdw$ as a function of $T$ for three different wave-vectors $\mathbf Q$ at coupling strength $\lambda_0 = 0.4$. See text for definitions of the different wave-vectors. Other parameters are $\omega_0 / \Ef = 0.1$, $n = 0.8$, and $L = 12$.}
 \label{fig:supp_Xcdw_vs_T}
\end{figure}

We focus on the case $\lambda_0 = 0.4$, which is roughly the largest coupling for which ME theory appears to be valid for the SC properties. 
  Figure \ref{fig:supp_Xcdw_vs_T} compares the temperature evolution of $\Xcdw({\mathbf Q})$ from DQMC and ME theory, for three different wave-vectors $\mathbf Q =(\pi, \pi)$, $(\pi, 2\pi / 3)$, and  $(\pi, 0)$.  These correspond, respectively, to the wave-vectors at which $\Xcdw$ computed from DQMC is maximal,  $\mathbf Q_{max}^{DQMC}=(\pi,\pi)$, at which the low $T$ ME expression for the same quantity is maximal, $\mathbf Q_{max}^{ME}\approx (\pi,2\pi/3)$, and at a typical non-maximal $\mathbf Q$.  
Note that while  $\mathbf Q_{max}^{DQMC}$ is $T$ independent, at elevated temperatures the  ME expression for $\Xcdw$ is maximal at $ (\pi, \pi)$ and  only shifts to  $\mathbf Q_{max}^{ME}$ below $T\approx t/16$. 
As is already implicit in Fig. \ref{fig:w0_renorm}, ME theory correctly predicts the SC properties of the system for $\lambda_0=0.4$.
 Indeed, as can already be seen in Fig. \ref{fig:w0_renorm}, it  correctly reproduces   $\Xcdw(\mathbf Q)$ for most values of $\mathbf Q$.  However, the ME expression produces unphysical peaks in $\Xcdw$ at a $\mathbf Q$ that is overly sensitive to the Fermi surface structure, and the peak value of $\Xcdw$ is not as strongly divergent with decreasing $T$ as the DQMC results.  Both trends are, of course, much more dramatic for $\lambda_0=0.5$, as can be inferred from Fig. \ref{fig:w0_renorm}.

%%%%%%%%%%%%%%%%%%%%%%%%%%%%%%%%%%%%%%%%%%%
%%%%%%%%%%%%%%%%%%%%%%%%%%%%%%%%%%%%%%%%%%%

\section{Adiabatic ratio $\wo/\Ef=1$}

In addition to the value of the adiabatic parameter $\wo/\Ef = 0.1$ discussed in the main text, we have also studied $\wo/\Ef = 1$. While this makes the nominal control parameter $\lambda_0 \wo/\Ef$ of ME theory larger, the larger phonon frequency also suppresses CDW ordering. The competition between an increase in the magnitude of vertex corrections and suppression of CDW ordering makes this an interesting regime in which to examine the validity of ME theory. In addition to the standard ME theory discussed in the main text, for this value of $\wo/\Ef$ we also compare with a version of the ME equations in which the \textit{bare} phonon propagator is used and only the electron self-energy is computed self-consistently. Because this approximation involves only the single rainbow diagram in Figure \ref{fig:Migdal_Equations} of the main text, we call this the ``rainbow approximation" (RA). This version of the approximation has often been used in the literature when assessing the validity of ME theory for the e-p problem \cite{PhysRevB.40.197,PhysRevLett.66.778}.

\begin{figure}[h]
\includegraphics[]{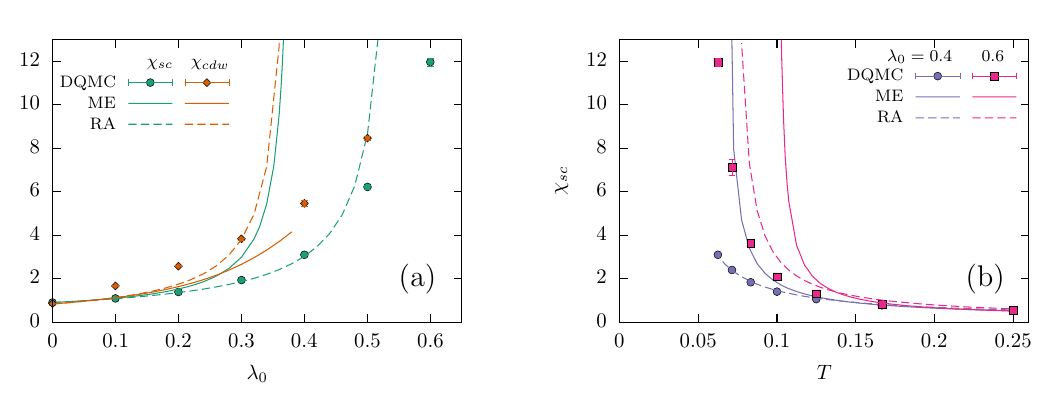}
\caption{(a) $\Xsc$ and $\Xcdw$ as a function of $\lambda_0$ for $\wo/\Ef = 1$, $\beta t=16$, $n=0.8$, and $L=12$. (b) $\Xsc$ as a function of $T$ for $\lambda_0 = 0.4$ and $0.6$, with other parameters the same as in panel (a). }
\label{fig:supp_Xsc_vs_lambda}
\end{figure}

In Figure \ref{fig:supp_Xsc_vs_lambda}(a) we plot the SC and CDW susceptibilities $\Xsc$ and $\Xcdw$ as a function of $\lambda_0$ for fixed parameters $\wo/\Ef = 1$, $\beta t= 16$, $n=0.8$, and $L=12$. We find the ME prediction for $\Xsc$ begins to deviate from the DQMC for $\lambda_0 \gtrsim 0.3$ and diverges shortly after, indicating a transition to a SC state. This is in contrast what was found for $\wo/\Ef = 0.1$, where the ME theory broke down for $\lambda_0 \approx 0.4$ (see main text) and the breakdown correlated with a sharp increase in the growth of CDW correlations. On the other hand, the RA agrees well with the measured $\Xsc$ up to $\lambda_0 \approx 0.5$. We find that neither the standard ME approximation nor the RA provide accurate estimates of $\Xcdw$ for $\lambda_0 \gtrsim 0.1$.

In Figure \ref{fig:supp_Xsc_vs_lambda}(b) we plot $\Xsc$ as a function of temperature for $\lambda_0 = 0.4$ and $0.6$, with all other parameters the same as in Fig. \ref{fig:supp_Xsc_vs_lambda}(a). For $\wo/\Ef = 0.1$ and $\lambda_0=0.4$, it was shown in the main text that ME theory captures $\Xsc$ accurately over the entire temperature range. However, for $\wo/\Ef = 1$ and $\lambda_0=0.4$, the ME prediction becomes inaccurate for $T\lesssim 0.1 t$. On the other hand, the RA is accurate across the entire temperature range. 
For $\lambda_0 = 0.6$ the ME prediction becomes inaccurate at a temperature $T\approx 0.15 t$, although the qualitative behavior is still correct. Based on the divergent behavior of $\Xsc$, the RA presumably predicts a SC transition temperature close to that predicted by the DQMC, although we have not attempted an accurate determination of $T_c$.

%%%%%%%%%%%%%%%%%%%%%%%%%%%%%%%%%%%%%%%%%%%
%%%%%%%%%%%%%%%%%%%%%%%%%%%%%%%%%%%%%%%%%%%

\end{document}